\documentclass[prd,preprintnumbers,nofootinbib,superscriptaddress]{revtex4} 

\usepackage{epsfig,graphicx}
\usepackage{dcolumn}
\usepackage{bm}
\usepackage{latexsym}
\usepackage{amsfonts}
\usepackage[usenames,dvipsnames]{color}
\usepackage{enumerate}
\usepackage{amssymb}
\usepackage{amsmath}

\newcommand{\Comment}[1]{{}}
\definecolor{MyDarkBlue}{rgb}{0.15,0.15,0.45}
\usepackage[linktocpage=true]{hyperref}
\hypersetup{
colorlinks=true,
citecolor=MyDarkBlue,
linkcolor=MyDarkBlue,
urlcolor=MyDarkBlue,
pdfsubject={hep-th}
}

\newcommand{\be}{\begin{equation}}
\newcommand{\ee}{\end{equation}}
\newcommand{\bea}{\begin{eqnarray}}
\newcommand{\eea}{\end{eqnarray}}
\newcommand{\beas}{\begin{eqnarray*}}
\newcommand{\eeas}{\end{eqnarray*}}
\newcommand{\nn}{\nonumber}

\begin{document}

\title{Cosmological perturbations of massive gravity coupled to DBI Galileons}

\author{Melinda Andrews}
\email{mgildner@sas.upenn.edu}
\affiliation{Center for Particle Cosmology, Department of Physics and Astronomy, University of Pennsylvania, Philadelphia, PA 19104}

\author{Kurt Hinterbichler}
\email{khinterbichler@perimeterinstitute.ca}
\affiliation{Perimeter Institute for Theoretical Physics, 31 Caroline St. N, Waterloo, Ontario, Canada, N2L 2Y5}

\author{James Stokes}
\email{stokesj@sas.upenn.edu}
\affiliation{Center for Particle Cosmology, Department of Physics and Astronomy, University of Pennsylvania, Philadelphia, PA 19104}

\author{Mark Trodden}
\email{trodden@physics.upenn.edu}
\affiliation{Center for Particle Cosmology, Department of Physics and Astronomy, University of Pennsylvania, Philadelphia, PA 19104}

\date{\today}

\begin{abstract}
Certain scalar fields with higher derivative interactions and novel classical and quantum mechanical properties -- the Galileons -- can be naturally covariantized by coupling to nonlinear massive gravity in such a way that their symmetries and number of degrees of freedom are unchanged.  We study the propagating degrees of freedom in these models around cosmologically interesting backgrounds. We identify the conditions necessary for such a theory to remain ghost free, and consider when tachyonic instabilities can be avoided.  We show that on the self-accelerating branch of solutions, the kinetic terms for the vector and scalar modes of the massive graviton vanish, as in the case of pure massive gravity.
\end{abstract}

\maketitle

\section{Introduction}
Over the past decade, attempts to find a consistent infrared modification of gravity have led to two seemingly distinct discoveries. The first is a consistent, ghost free nonlinear realization of massive gravity, known as dRGT massive gravity~\cite{deRham:2010ik,deRham:2010kj} (see \cite{Hinterbichler:2011tt} for a review). The second is a class of intriguing scalar field theories -- the Galileons~\cite{Nicolis:2008in} (see \cite{deRham:2012az} for a review) -- with novel classical and quantum properties that can be traced to their nonlinear derivative interactions. There is at least one connection between these ideas, in that the Galileon interactions govern the longitudinal degree of freedom of a ghost-free massive graviton in the decoupling limit \cite{deRham:2010ik}. Furthermore, if one is to consider covariantizing Galileons, while preserving second order equations of motion and the special symmetries of those theories, then it is natural to couple not to General Relativity (GR) as in \cite{Deffayet:2009wt,Deffayet:2009mn} (which breaks the Galileon symmetry), but rather to massive gravity itself \cite{Gabadadze:2012tr}.  

Besides the natural question of how best to covariantize the Galileons, there are other reasons for considering the theory of massive gravity Galileons. The reason that theories of massive gravity provide an opportunity to address the problem of late-time cosmic acceleration without a cosmological constant is that they often admit self-accelerating vacuum solutions. In addition, a light graviton mass is technically natural, in contrast to a cosmological constant which receives large radiative corrections. However, although the dRGT theory possesses a self-accelerating solution with negatively curved spatial slices~\cite{Gumrukcuoglu:2011ew}, the study of fluctuations on top of this background has shown that the kinetic terms for the vector and scalar perturbations vanish~\cite{Gumrukcuoglu:2011zh}. The vanishing of these terms can be remedied by departing from isotropic and homogeneous cosmologies \cite{Gumrukcuoglu:2012aa,DeFelice:2013awa} or by introducing new degrees of freedom. There are many ways to achieve the latter option, and several possibilities have been explored in the so-called quasi-dilaton~\cite{D'Amico:2012zv,Haghani:2013eya,Gumrukcuoglu:2013nza,D'Amico:2013kya} and mass-varying extensions of dRGT \cite{D'Amico:2011jj,Huang:2012pe,Gumrukcuoglu:2013nza}.  (See \cite{DeFelice:2013bxa} for a review of these aspects of cosmological solutions in various versions of massive gravity.)

In this paper we perform a study of cosmological perturbations for the natural extension of dRGT introduced in \cite{Gabadadze:2012tr} -- massive gravity Galileons -- in which Dirac-Born-Infeld (DBI) scalars couple to the massive graviton in such a way that the scalars possess generalized Galileon shift symmetries~\cite{Gabadadze:2012tr}. This theory has been shown to be ghost free in the decoupling limit~\cite{Gabadadze:2012tr} and in full generality using the vielbein formalism~\cite{Andrews:2013ora} (see also \cite{Kluson:2013hoa}), and admits a self-accelerating branch that is a generalization of that discovered for massive gravity itself \cite{Hinterbichler:2013dv}. We study fluctuations around the self-accelerating branch and show that the kinetic terms for the scalar and vector modes vanish, just as they do in pure dRGT theory.

\section{Massive Gravity-Galileon Theory}

The construction of massive gravity coupled to Galileons is carried out using an extension of the probe brane approach~\cite{deRham:2010eu,Hinterbichler:2010xn,Goon:2011xf,Goon:2011qf,Goon:2011uw,Burrage:2011bt} for constructing general Galileon models and the bi-metric approach for constructing the dRGT nonlinear massive gravity theory \cite{Hassan:2011tf,Hassan:2011zd}. We introduce a physical metric $g_{\mu\nu}$ and a second, induced metric $\bar{g}_{\mu\nu}$ which is the pull-back through a dynamical embedding $\phi^A(x)$ into a five-dimensional Minkowski space with metric $\eta_{AB} = {\rm diag}(-1,1,1,1,1)$,
\begin{equation}
	\bar{g}_{\mu\nu} = \eta_{AB} \partial_\mu \phi^A \partial_\nu \phi^B \ .
\end{equation}
 
The action contains three kinds of terms:
\begin{align}\label{actionm}
S&\!=\! S_{\rm EH}[g]+\!S_{\rm mix}[g,\bar g]+\!S_{\rm Galileon}[\bar g] \ .
\end{align}
The first part $S_{\rm EH}[g]$ is the Einstein-Hilbert action for $g_{\mu\nu}$
\be S_{\rm EH}[g]=\frac{M_{\rm P}^{2}}{2} \int {\rm d}^4 x\, \sqrt{-g}\,R[g] \ .
\ee
The second part is the action mixing the two metrics,
\begin{align}\label{mixaction}
S_{\rm mix}[g,\bar g]&= M_{\rm P}^{2}m^{2}\int {\rm d}^4 x \sqrt{-g} (\mathcal{L}_2 + \alpha_3 \mathcal{L}_3 + \alpha_4 \mathcal{L}_4)\ ,
\end{align}
where
\begin{eqnarray}
 {\cal L}_2 & = & \frac{1}{2}
  \left([{\cal K}]^2-[{\cal K}^2]\right)\,, \nonumber\\
 {\cal L}_3 & = & \frac{1}{6}
  \left([{\cal K}]^3-3[{\cal K}][{\cal K}^2]+2[{\cal K}^3]\right), 
  \nonumber\\
 {\cal L}_4 & = & \frac{1}{24}
  \left([{\cal K}]^4-6[{\cal K}]^2[{\cal K}^2]+3[{\cal K}^2]^2
   +8[{\cal K}][{\cal K}^3]-6[{\cal K}^4]\right)\,, \nn
\end{eqnarray}
and where the brackets are traces of powers of the matrix ${\cal K}^\mu_{\ \nu} = \delta^\mu_{\ \nu} - \sqrt{g^{\mu\sigma}\bar{g}_{\sigma\nu}}$. The final part is the DBI Galileon action $S_{\rm Galileon}[\bar g]$ consisting of the Lovelock invariants constructed from $\bar g$, and their boundary terms (see \cite{deRham:2010eu,Goon:2010xh,Goon:2011qf} and Sec. IV.B of \cite{Hinterbichler:2010xn}; we use normalizations consistent with \cite{Goon:2010xh,Goon:2011qf}),
\begin{equation}
S_{\rm Galileon} = \Lambda^4\int {\rm d}^4x \sqrt{-\bar{g}} \left\{ -a_2 + {a_3\over \Lambda} [\bar{K}] - {a_4 \over \Lambda^{2}} \left([{\bar K}]^2-[{\bar K}^2]\right) +{ a_5 \over \Lambda^{3}}  \left([{\bar K}]^3-3[{\bar K}][{\bar K}^2]+2[{\bar K}^3]\right)\right\},\label{galaction}
\end{equation}
where $\bar K_{\mu\nu}$ is the extrinsic curvature of the brane embedding $\phi^A(x)$ into the flat five-dimensional Minkowski space and indices are raised with $\bar g^{\mu\nu}$ (since the bulk is flat, we may use Gauss-Codazzi to eliminate all intrinsic curvatures in favor of extrinsic curvatures).

Note that we have set the cosmological constant and a possible tadpole term in $S_{\rm Galileon}$ to zero.  This ensures the existence of a flat space solution with constant $\pi$.
Restoring these terms does not change our essential conclusion.

\section{Background Cosmology and Self Accelerating Solutions}

For our purposes, we take a Friedmann-Robertson-Walker (FRW) ansatz for the physical metric
\begin{equation}
	\mathrm{d} s^2 = - N^2(t)\mathrm{d}t^2 + a^2(t)\Omega_{ij} \mathrm{d} x^i \mathrm{d}x^j~, \quad \Omega_{ij} = \delta_{ij} + \frac{\kappa}{1-\kappa r^2}x^i x^j \ ,
\end{equation}
where $\kappa<0$ is the spatial curvature.  

As shown in \cite{Hinterbichler:2013dv}, this model does not admit non-trivial cosmological solutions for a flat FRW ansatz with a homogeneous fiducial metric, just as pure dRGT massive gravity does not \cite{D'Amico:2011jj}, and there are no solutions for $\kappa>0$ since the fiducial Minkowski metric cannot be foliated by closed slices.  (There are, however, known solutions to pure dRGT massive gravity with FRW physical metric and and inhomogeneous St\"uckelberg sector \cite{D'Amico:2011jj,Gratia:2012wt}, that is, solutions where the physical metric is FRW but the fiducial metric is not also FRW in the same coordinates.) 

The embedding (the St\"uckelbergs) are chosen so that the fiducial metric has the symmetries of an open FRW spacetime \cite{Gumrukcuoglu:2011ew},
\begin{equation}
	\phi^0 = f(t)\sqrt{1 - \kappa \vec{x}^2}, \quad \phi^i = \sqrt{-\kappa}f(t)x^i, \quad \phi^5 \equiv \pi(t) \ .
\end{equation}
where $f(t)$ plays the role of a St\"uckelberg field which restores time reparametrization invariance.
The induced metric then takes the form
\begin{equation}
	{\bar g}_{\mu\nu}dx^\mu dx^\nu
		= \left(- \dot{f}(t)^2 +\dot\pi(t)^2\right)dt^2-\kappa f(t)^2 \Omega_{ij}(\vec x)dx^i dx^j.
\end{equation}
Note that we can obtain the extended massive gravity mass terms from the dRGT mass terms by replacing ${\bar g}_{\mu\nu}$ with ${\bar g}_{\mu\nu} + \partial_\mu \pi \partial_\nu \pi$.

This ansatz leads to the mini-superspace action 
\begin{align}
S_{\rm EH}
		& = 3 M_{\rm P}^2 \int {\rm d}t\,\left[ - \frac{\dot{a}^2a}{N} + \kappa N a\right] \ ,\\
	S_{\rm mix}
		& = 3 M_{\rm P}^2 \int{\rm d}t \, m^2 \left[NF(a,f) - G(a,f)\sqrt{\dot{f}^2 - \dot{\pi}^2} \right] \ , \\
	S_{\rm Galileon}
		& = \Lambda^4 \int{\rm d}t \, (\sqrt{-\kappa})^3 \Bigg[ -a_2 f^3 \sqrt{\dot{f}^2-\dot{\pi}^2} + \frac{a_3}	{\Lambda} f^2 \left( 3 \dot{\pi} + f \frac{\dot{f} \ddot{\pi} - \ddot{f} \dot{\pi }}{\dot{f}^2-\dot{\pi }^2} \right) \nonumber \\
		& \quad -6 \frac {a_ 4} {\Lambda^2} f \frac{\dot{\pi}}{\sqrt{\dot{f}^2 - \dot{\pi}^2}} \left( \dot{\pi} +  f \frac{\dot{f} \ddot{\pi} - \ddot{f}\dot{\pi}}{\dot{f}^2 - \dot{\pi}^2} \right)
+ 6\frac {a_5} {\Lambda^3} \frac{\dot{\pi}^2}{\dot{f}^2 - \dot{\pi}^2} \left( \dot{\pi} + 3 f \frac{\dot{f}\ddot{\pi} - \ddot{f}\dot{\pi}} {\dot{f}^2 - \dot{\pi}^2} \right) \Bigg] \ ,
\end{align}
where
\begin{align}
	F(a,f)
		& = a(a-\sqrt{-\kappa}f)(2a-\sqrt{-\kappa}f) 
		 +  \frac{\alpha_3}{3}(a-\sqrt{-\kappa}f)^2(4a-\sqrt{-\kappa}f) + \frac{\alpha_4}{3}(a-\sqrt{-\kappa}f)^3 \ , \label{Fofaf} \\ \nn
	G(a,f)
		& = a^2(a-\sqrt{-\kappa}f)+\alpha_3 a(a-\sqrt{-\kappa}f)^2  + \frac{\alpha_4}{3}(a-\sqrt{-\kappa}f)^3 \ . \label{Gofaf}
\end{align}

Varying with respect to $N$, we obtain the Friedmann equation,
\begin{equation}\label{DBINk}
	{H^2\over N^2} + {\kappa \over a^2} + m^2 {F(a,f)\over a^3}  = 0 \ .
\end{equation}

The equations obtained by varying the action with respect to $f$ and $\pi$, respectively, are
\begin{align}\label{DBIfequationk}
{\delta S\over \delta f} &= 3M_{\rm P}^2 m^2 \frac{\partial G}{\partial a} \left(\frac{\dot{a}}{\dot{f}} \sqrt{\dot{f}^2-\dot{\pi}^2} - N \sqrt{-\kappa}\right) + \frac{\dot{\pi}}{\dot{f}} \dot{\Pi} = 0 \ ,
\\ \label{DBIpiequationk}
\frac{\delta S}{\delta\pi} &= -\dot{\Pi} = 0 \ ,
\end{align}
where we have defined the quantity
\begin{align}
\Pi = &\left(3M_{\rm P}^2 m^2 G+a_2 \Lambda^4 \left(\sqrt{-\kappa}f\right)^3 \right) \frac{\dot{\pi}}{\sqrt{\dot{f}^2-\dot{\pi}^2}}
-3 a_3 \Lambda^3 \left(\sqrt{-\kappa}\right)^3 f^2 \left(\frac{\dot{\pi}}{\sqrt{\dot{f}^2-\dot{\pi}^2}}\right)^2 
\nonumber \\
&+6 a_4 \Lambda^2 \left(\sqrt{-\kappa}\right)^3 f \left(\frac{\dot{\pi}}{\sqrt{\dot{f}^2-\dot{\pi}^2}}\right)^3  
-6 a_5 \Lambda \left(\sqrt{-\kappa}\right)^3 \left(\frac{\dot{\pi}}{\sqrt{\dot{f}^2-\dot{\pi}^2}}\right)^4 
\ .
\end{align}
The acceleration equation obtained by varying with respect to $a$ is redundant, due to the time reparametrization invariance of the action.

In contrast to GR, these equations enforce a constraint: the combination $\dot f{\delta S\over \delta f}+\dot \pi{\delta S\over \delta \pi}$ becomes
\begin{equation}\label{DBIconstraintk}
	{\partial G(a,f)\over \partial a}\left(\dot a\sqrt{\dot f^2-\dot\pi^2}-\sqrt{-\kappa}N \dot f\right)= 0 \ ,
\end{equation}
the analogue of which for pure massive gravity is responsible for the well-known absence of flat FRW solutions in that theory.

There exist two branches of solutions depending on whether the constraint equation is solved by setting ${\partial G\over\partial a}=0$ or instead by setting $\dot a\sqrt{\dot f^2-\dot\pi^2}-\sqrt{-\kappa}N\dot f = 0$. In this work we shall focus on the former choice, since this corresponds to de Sitter space -- the self-accelerating branch of the theory \cite{Hinterbichler:2013dv}, in which the metric takes the same form as the self-accelerating solution of the original massive gravity theory.  Defining 
\be X\equiv {\sqrt{-\kappa}f\over a}~,\label{Xdef}\ee
we find an algebraic equation for $f$ that can be written in the form $J_\phi = 0$, where
\begin{equation}\label{Jphidef}
 J_{\phi} \equiv 3-2X+\alpha_3(1-X)(3-X)+\alpha_4(1-X)^2~.
\end{equation}
The solutions read
\begin{equation}
 f(t) = \frac{1}{\sqrt{-\kappa}}X_{\pm}a(t)\,,
\qquad
X_{\pm} \equiv \frac{1+2\,\alpha_3+\alpha_4
 \pm\sqrt{1+\alpha_3+\alpha_3^2-\alpha_4}}
 {\alpha_3+\alpha_4} \ .
\label{eq:fsolcosmo}
\end{equation}
These are the same self-accelerated solutions that were found in pure massive gravity~\cite{Gumrukcuoglu:2011ew}.  The solution for the extra Galileon field $\pi$ can then be determined by solving \eqref{DBIpiequationk}.

\section{Perturbations}
We now turn to the primary issue addressed in this paper -- the behavior of perturbations around this background cosmological solution. To obtain the quadratic action for perturbations, we work in unitary gauge for the St\"uckelberg fields $\phi^0$ and $\phi^i$ and expand the metric and $\pi$ fields to second order in fluctuations around the background. We write the metric as $g_{\mu\nu}=g^{(0)}_{\mu\nu} +\delta g_{\mu\nu}$, with
\begin{equation}
	\delta g_{\mu\nu} = 
	\begin{pmatrix}
	-2N^2 \Phi & Na B_i \\
	Na B_j & a^2 h_{ij}
	\end{pmatrix}. \label{gpert}
\end{equation}
Here, $\Phi,B_i$ and $h_{ij}$ are the small perturbations, $N$ and $a$ are the background lapse and scale factor, and we henceforth raise and lower latin indices with respect to $\Omega_{ij}$.

The vector perturbation $B_i$ can be decomposed into transverse and longitudinal components via
\begin{equation}\label{Bdecomp}
	B_i = B^{T}_i + \partial_i B \ , \ \ \   D^iB^{T}_i=0,
\end{equation}
where  $D_i$ denotes the covariant derivative with respect to $\Omega_{ij}$.   The tensor perturbations $h_{ij}$ decompose into a transverse traceless component $h^{TT}_{ij}$, a transverse vector $E^T_i$, a longitudinal component $E$, and a trace $\Psi$ as follows:
\begin{equation}\label{hdecomp}
 h_{ij} = 2 \Psi \Omega_{ij} +\left(D_iD_j - \frac{1}{3}\Omega_{ij}\triangle\right)E + \frac{1}{2}(D_i E^T_j + D_j E^T_i) + h^{TT}_{ij} \ ,
\end{equation}
 where $\triangle \equiv D^iD_i$, and the transverse traceless conditions read
\be 
D^ih_{ij}^{TT} = h^{TT\;i}_i=0,\ \ \ D^iE^{T}_i=0 \ .
\ee

We denote the remaining dynamical scalar field -- the Galileon perturbation -- by $\tau$, via
\begin{equation}
	\phi^5 = \pi + \tau \ . \label{galexpansiondef}
\end{equation}

\subsection{Preliminaries}

Before writing the full quadratic actions for the various perturbations, we first write some intermediate expressions obtained from the expansions of the mass terms \eqref{mixaction}.  This will serve to highlight the manner in which the kinetic terms vanish, and illustrate the similarities with pure dRGT.

For convenience, we introduce the quantities
\begin{equation}\label{srdefs}
	s = \sqrt{1 - (\dot{\pi}/{\dot f})^2}~, \quad r = \frac{{\dot f}a}{N\sqrt{-\kappa}f} \ ,
\end{equation} 
and we will continue to use $J_\phi$ to denote the quantity \eqref{Jphidef} which vanishes on the equations of motion.

Expanding the mass term to linear order in the fluctuations yields
\begin{equation}
 S_{\rm mix} = S^{(0)}_{\rm mix} + 
  \int dx^4Na^3\sqrt{\Omega}
  \left[-\left(\Phi+\frac{1}{2}h\right)\rho_g
   +\frac{1}{2}M_{\rm P}^2m_g^2(1-rs)XhJ_{\phi} + M_{\rm P}^2m_g^2(r\dot{\pi}/{\dot f}^2s)Y\dot{\tau}\right] \ ,
\end{equation}
where we have defined
\begin{align}
 \rho_g 
 	& \equiv -M_{Pl}^2m_g^2(1-X)
  \left[3(2-X)+\alpha_3(1-X)(4-X)+\alpha_4(1-X)^2\right]  ,\\
  Y
  	& \equiv X(1-X)\left[3+3\alpha_3(1-X)+\alpha_4(1-X)^2\right] \ .
\end{align}
When the background equation of motion for the St\"uckelberg fields are satisfied, the terms linear in the metric match the corresponding terms of pure massive gravity. This suggests that we follow the massive gravity analysis of \cite{Gumrukcuoglu:2011zh} and define
\begin{equation}
 \tilde{S}_{\rm mix}[g_{\mu\nu},\tau] \equiv 
  S_{\rm mix}[g_{\mu\nu},\tau] + 
  \int {\rm d}^4x\sqrt{-g}\rho_g\equiv M_{\rm P}^2 m_g^2 \int d^4 x N a^3 \sqrt{\Omega} \tilde{\mathcal{L}}_{\rm mix} \ .
\end{equation}
Expanding to second order in perturbations we have,
\begin{align}
\tilde{\mathcal{L}}_{\rm mix}^{(0)}  = & 
  -rs Y, \\
 \tilde{\mathcal{L}}_{\rm mix}^{(1)}  = & 3(1-rs)XJ_{\phi} \Psi  + (r\dot{\pi}/{\dot f}^2s)Y\dot{\tau}, \\
\tilde{\mathcal{L}}_{\rm mix}^{(2)} = &\frac{1}{2} \frac{r}{s} \frac{1}{{\dot f}^2 s^2} Y \dot{\tau}^2 + \frac{1}{2} \left(6\Phi\Psi + \frac{B^T_i B^{Ti}}{1+rs} \right) X J_\phi + 3 \frac{r}{s} \frac{\dot{\pi}}{{\dot f}^2} X J_\phi \dot{\tau} \Psi  \nn \\
&+ \frac{\dot{\pi}}{\sqrt{-\kappa}{\dot f}f} \left(\frac{r}{1+rs}\right) X J_\phi \tau\triangle B  - \frac{1}{2\kappa f^2} \left[\left(\frac{1-r^2}{1+rs}\right) X J_\phi + \frac{r}{s} Y \right] \tau \triangle \tau \nn  \\
&+ \frac{1}{8} (1-rs) \left( 12\Psi^2 - 2 h^{TT}_{ij}h^{TTij} +E^T_j \triangle E^{Tj} \right) X J_\phi \nn  \\
& + \frac{1}{8} m_g^{-2} M_{GW}^2 \left( 24\Psi^2 - h^{TT}_{ij}h^{TTij} + \frac{1}{2}E^T_j \triangle E^{Tj}  \right), \label{L2lag}
\end{align}
where we have defined a quantity which will turn out to be the graviton mass term:
\begin{equation}
 m_g^{-2}M_{GW}^2 \equiv 
  XJ_{\phi} + (1-rs)X^2\left[1+\alpha_3(2-X)+\alpha_4(1-X)\right] \ .
\end{equation}
Here we have not imposed any equations of motion on the background.
We note that all of the terms in \eqref{L2lag} which depend upon $\Phi$ or $B_i$ are proportional to $J_\phi$, and therefore vanish on the de Sitter self-accelerating branch, on which $J_\phi=0$.  As we will see, this implies the vanishing of the graviton scalar and vector kinetic terms on this background.

\subsection{Tensor perturbations}

We now write the full second order action obtained from expanding \eqref{actionm} and decomposing the perturbations according to \eqref{gpert} and \eqref{Bdecomp}, \eqref{hdecomp}.

The tensor perturbations take the same form as in pure massive gravity, but with a different time-dependent mass term,
\begin{equation}
 S^{(2)}_{\rm tensor} = 
  \frac{M_{\rm P}^2}{8}\int {\rm d}^4x \sqrt{\Omega}\, Na^3
  \left[\frac{1}{N^2}\dot{h}^{TTij}\dot{h}^{TT}_{ij}
   + \frac{1}{a^2}h^{TTij}(\triangle-2\kappa)h^{TT}_{ij}
   -M_{GW}^2 h^{TTij}h^{TT}_{ij}\right],
\end{equation}
where $M_{GW}^2$, in terms of the definitions \eqref{srdefs}, \eqref{Xdef} made above, takes the following value on the de Sitter self-accelerating branch,\begin{equation}
	M_{GW}^2 = \pm (rs - 1) m_g^2 X_\pm^2\sqrt{1+\alpha_3 + \alpha_3^2 - \alpha_4}~.
\end{equation}

As in pure massive gravity, the tensor perturbation maintains the correct sign for both the kinetic and gradient terms. However, the new mass term implies a  more complicated region of parameter space in which the tensors are non-tachyonic, $M_{GW}^2>0$ (the sign of the mass term is given by the sign of $\pm (rs - 1)$).  Note, however, that even if this term is negative, so that we have a tachyonic instability, then barring any fine tuning such instabilities are of order the Hubble scale if we have chosen $m\sim H$ to ensure late time acceleration of the correct magnitude. This agrees qualitatively with the result found in pure massive gravity.

\subsection{Vector perturbations}
Since the vector field $B_i^T$ obtained from $\delta g_{0i}$ does not appear in the Lagrangian with any time derivatives, it can be eliminated as an auxiliary field. Leaving the background fields arbitrary for the moment, we find the solution
\begin{equation}
	B^T_i = \frac{a (1+r s) \left(-\triangle-2 \kappa \right)}{2 \left[(1+r s)(-\triangle-2\kappa)+2 a^2 J_\phi m^2 X\right] N} \dot{E}^T_i.
\end{equation}
This matches the result of pure dRGT theory $B^T_i = \frac{a}{2N} \dot{E}^T_i$ when the St\"uckelberg equation of motion for the de Sitter self-accelerating branch is imposed, $J_\phi=0$. It is instructive, however, to leave the backgrounds arbitrary so that we can explicitly see the kinetic term vanish. Substituting the general expression for the non-dynamical vector we obtain
\begin{equation}
S_{\rm vector}^{(2)} = \frac{M_{\rm P}^2}{8} \int  d^4x \sqrt{\Omega}\,  a^3 N  \left\{ \mathcal{T}_V (\dot{E}_i^T)^2 - \left[ \frac{1}{2} M_{GW}^2 (-\triangle - 2\kappa) + J_\phi k^2 m^2(1-rs) \right](E_i^T)^2\right\}~,
\end{equation}
where
\begin{equation}
	\mathcal{T}_V =\frac{a^2 J_\phi m^2 X \left(-\triangle-2 \kappa \right)}{\left[(1+r s)(-\triangle - 2\kappa)+2 a^2 J_\phi m^2 X\right] N^2} \ .
\end{equation}
The vanishing of the vector kinetic terms is now obvious on the de Sitter self-accelerating branch where $J_\phi = 0$.  The vector Lagrangian has the same form as pure dRGT theory, only with a different time-dependent mass,
\begin{equation}
S_{\rm vector}^{(2)} = -\frac{M_{\rm P}^2}{16} \int d^4x \sqrt{\Omega}\, a^3 N  M_{GW}^2 (-\triangle - 2\kappa) (E_i^T)^2 .
\end{equation}

\subsection{Scalar perturbations}
The analysis of the scalar perturbations simplifies considerably on the de Sitter self-accelerating branch since all the terms mixing scalar degrees of freedom from the graviton with the fluctuation of the Galileon vanish when $J_\phi = 0$, as can be seen from the expression \eqref{L2lag}. The scalars $\Phi$ and $B$ coming from perturbations of $\delta g_{00}$ and $\delta g_{0i}$ appear without time derivatives and we may eliminate them as auxiliary fields.  We obtain (this time imposing the self-accelerating background equation of motion $J_\phi = 0$)
\begin{align}
 	\Phi
		& = \frac{\kappa\triangle}{6a^2 H^2} E -  \frac{\triangle}{6HN}\dot{E} - \frac{\kappa}{a^2 H^2}\Psi + \frac{1}{HN}\dot{\Psi} \\
	B
		& =  \frac{\triangle}{6aH} E + \frac{a}{2N}\dot{E} - \frac{1}{aH}\Psi
\end{align}
which are the same as in pure dRGT theory. The calculation of the graviton scalar quadratic action mirrors the dRGT case and we find that the kinetic terms vanish and the action once again has the same form as pure dRGT, only with a modified time-dependent mass,
\begin{equation}
	S_{\rm scalar}^{(2)}
		= \frac{M_{\rm P}^2}{2} \int d^4x\sqrt{\Omega}\,  a^3N  \left(6 M_{GW}^2 \Psi^2+  \frac{1}{6} M_{GW}^2 \triangle (-\triangle - 3\kappa) E^2 \right).
\end{equation}

We now turn to the expansion of the Galileon action \eqref{galaction}, using \eqref{galexpansiondef}. We start by expanding the lowest Galileon, the DBI term (the one proportional to $a_2$ in \eqref{galaction}) to quadratic order in $\tau$.  We obtain $S_{\rm DBI} = -a_2\Lambda^4\int d^4 x Na^3\sqrt{\Omega}\mathcal{L}_{\rm DBI}$, where
\begin{align}
	\mathcal{L}_{\rm DBI}^{(0)}
		& = rs X^4  ~, \\
	\mathcal{L}_{\rm DBI}^{(1)}
		& = -\left(\frac{r}{s} \frac{\dot{\pi}}{{\dot f}^2}\dot{\tau} \right) X^4~, \\
	\mathcal{L}_{\rm DBI}^{(2)}
		& = -\frac{1}{2}\frac{r}{s} \left(\frac{1}{{\dot f}^2s^2}\dot{\tau}^2+\frac{1}{\kappa f^2} \tau \triangle \tau\right) X^4 \ .
\end{align}
From $\mathcal{L}_{\rm DBI}^{(2)}$ we see that the effect of including the DBI Lagrangian is to shift $Y \to Y + (a_2\Lambda^4/m^2M_{\rm P}^2) X^4$ in the quadratic action \eqref{L2lag}. Note that on the de Sitter self-accelerating branch, where $J_\phi = 0$, this is equivalent to shifting the brane tension by
\begin{equation}
	\Lambda^4 \to \tilde{\Lambda}^4 = \Lambda^4 +  {m^2 M_{\rm P}^2 \over a_2}\frac{Y_\pm}{X_\pm^4} ~.
\end{equation}
We therefore see that on the self-accelerating de Sitter branch, the Galileon has the correct-sign kinetic term provided
\be
\frac{m^2 M_{\rm P}^2}{a_2\Lambda^4} \frac{Y_\pm}{X_\pm^4} > -1 \ .
\ee
It is clear that this constraint can always be satisfied by choosing $a_2\Lambda^4$ appropriately large. Note that the background St\"uckelberg and Galileon fields do not lead to any simplification for the DBI quadratic action. 

The higher Galileon terms in \eqref{galaction} can be similarly expanded to quadratic order. After imposing the background equation for the Stueckelberg/Galileon and its time derivatives, we obtain $S_{\rm Galileon} = \Lambda^4\int d^4 x Na^3\sqrt{\Omega}\mathcal{L}_{\rm Galileon}$, where
\begin{align}
	\mathcal{L}_{\rm Galileon}^{(2)}
		& = - \frac{r}{\dot{f}^2s^3}\left[-\frac{1}{2}a_2 + 3\frac{a_3}{\Lambda}\left(\frac{\dot{\pi}}{sf\dot{f}}\right)-9\frac{a_4}{\Lambda^2}\left(\frac{\dot{\pi}}{sf\dot{f}}\right)^2 + 12\frac{a_5}{\Lambda^3}\left(\frac{\dot{\pi}}{sf\dot{f}}\right)^3\right]X^4\dot{\tau}^2 \nonumber \\  
		& \quad 
		- \frac{r}{s}\frac{1}{\kappa f^2}\left[ 
		 \frac{1}{2} a_2 
		+ \frac{a_3}{\Lambda}\frac{\dot{\pi}}{sf\dot{f}} 
		- \frac{a_4}{\Lambda^2}\frac{3\dot{\pi}^4 + 11 \dot{f}^2\dot{\pi}^2 - 2\dot{f}^4}{s^2f^2\dot{f}^4}
		+
		 6\frac{a_5}{\Lambda^3}\frac{3\dot{\pi}^4+2\dot{f}^2\dot{\pi}^2 + 2 \dot{f}^4}{s^3f^3\dot{f}^5}\dot{\pi} \right]X^4 \tau \triangle \tau ~.
\end{align}
The conditions for stability can now be read off by requiring that these kinetic terms have the correct sign.

\section{Conclusions}
We have examined the nature of cosmological perturbations around the self-accelerating branch of the massive gravity Galileon theory of \cite{Gabadadze:2012tr}, in which the Galileon fields couple covariantly to massive gravity while simultaneously retaining both their symmetry properties and their second order equations of motion. This construction provides a more general framework within which we may ask self-consistent questions about the implications of both massive gravity and Galileon models. Such an approach is important both for the wider goal of understanding the general implications of modified gravity models, and also for probing the extent to which particular features of massive gravity are peculiar to the specific restrictive structure of that theory.

One of the more striking results of dRGT massive gravity is that the kinetic terms for both vector and scalar perturbations vanish around the phenomenologically interesting self-accelerating branch of the theory. The main result of our analysis is that the vanishing of these kinetic terms is preserved around the analogous de Sitter branch in the more general class of theories, suggesting that this is a generic result tied to the existence of self-accelerating homogeneous solutions in theories with this general structure. This is partly upheld by ghost-free bimetric gravity, for which there are two branches of FRW solutions: one for which the ratio of Hubble constants for the two metrics is constant, and one for which it is non-constant.  The first branch is the analogue of the solution studied here, and exhibits the same vanishing of kinetic terms for one vector and one scalar degree of freedom; on the second branch, however, all 7 degrees of freedom are dynamical \cite{Comelli:2012db}. Furthermore, we have verified that the tensor perturbations are ghost-free, and that while the details of the analysis of their mass terms differs from that in pure massive gravity, any tachyonic modes are similarly unstable on Hubble timescales. 

An obvious extension of this work is to study the behavior of the same perturbations around the other branch of cosmological solutions identified in \cite{Hinterbichler:2013dv} (this work is underway).  One may also wish to search for the equivalent to the pure dRGT massive gravity solutions which display flat, approximately FRW cosmological solutions at the cost of an inhomogeneous St\"uckelberg sector as in \cite{D'Amico:2011jj,Gratia:2012wt}, about which the kinetic terms for scalar perturbations no longer vanish \cite{Wyman:2012iw,Koyama:2011wx}.  It will also be interesting to ask whether the fluctuations of the Galileon around these accelerating solutions can be kept subluminal, in contrast to the situation around flat space \cite{Goon:2010xh}.

\vspace{5mm}
{\bf Acknowledgments:}
We would like to thank Shinji Mukohyama for inviting this article, and for his patience with us during our delays in submitting it, and A. Emir G\"umr\"uk\c{c}\"uo\u{g}lu for pointing out an error in our previous mini-superspace expressions for the higher galileons. Work at the University of Pennsylvania was supported in part by the US Department of Energy, and NASA ATP grant NNX11AI95G. Research at Perimeter Institute is supported by the Government of Canada through Industry Canada and by the Province of Ontario through the Ministry of Economic Development and Innovation. The work of KH was made possible in part through the support of a grant from the John Templeton Foundation. The opinions expressed in this publication are those of the authors and do not necessarily reflect the views of the John Templeton Foundation.


\begin{thebibliography}{99}


\bibitem{deRham:2010ik} 
  C.~de Rham and G.~Gabadadze,
  Phys.\ Rev.\ D {\bf 82}, 044020 (2010)
  [arXiv:1007.0443 [hep-th]].


\bibitem{deRham:2010kj} 
  C.~de Rham, G.~Gabadadze and A.~J.~Tolley,
  Phys.\ Rev.\ Lett.\  {\bf 106}, 231101 (2011)
  [arXiv:1011.1232 [hep-th]].

\bibitem{Hinterbichler:2011tt} 
  K.~Hinterbichler,
  Rev.\ Mod.\ Phys.\  {\bf 84}, 671 (2012)
  [arXiv:1105.3735 [hep-th]].

\bibitem{Nicolis:2008in} 
  A.~Nicolis, R.~Rattazzi and E.~Trincherini,
  Phys.\ Rev.\ D {\bf 79}, 064036 (2009)
  [arXiv:0811.2197 [hep-th]].
  
\bibitem{deRham:2012az} 
  C.~de Rham,
  Comptes Rendus Physique {\bf 13}, 666 (2012)
  [arXiv:1204.5492 [astro-ph.CO]].

\bibitem{Deffayet:2009wt} 
  C.~Deffayet, G.~Esposito-Farese and A.~Vikman,
  Phys.\ Rev.\ D {\bf 79}, 084003 (2009)
  [arXiv:0901.1314 [hep-th]].


\bibitem{Deffayet:2009mn} 
  C.~Deffayet, S.~Deser and G.~Esposito-Farese,
  Phys.\ Rev.\ D {\bf 80}, 064015 (2009)
  [arXiv:0906.1967 [gr-qc]].

\bibitem{deRham:2010eu} 
  C.~de Rham and A.~J.~Tolley,
  JCAP {\bf 1005}, 015 (2010)
  [arXiv:1003.5917 [hep-th]].


\bibitem{Goon:2011xf} 
  G.~Goon, K.~Hinterbichler and M.~Trodden,
  JCAP {\bf 1112}, 004 (2011)
  [arXiv:1109.3450 [hep-th]].


\bibitem{Hinterbichler:2010xn} 
  K.~Hinterbichler, M.~Trodden and D.~Wesley,
  Phys.\ Rev.\ D {\bf 82}, 124018 (2010)
  [arXiv:1008.1305 [hep-th]].


\bibitem{Goon:2011qf} 
  G.~Goon, K.~Hinterbichler and M.~Trodden,
  JCAP {\bf 1107}, 017 (2011)
  [arXiv:1103.5745 [hep-th]].


\bibitem{Goon:2011uw} 
  G.~Goon, K.~Hinterbichler and M.~Trodden,
  Phys.\ Rev.\ Lett.\  {\bf 106}, 231102 (2011)
  [arXiv:1103.6029 [hep-th]].


\bibitem{Burrage:2011bt} 
  C.~Burrage, C.~de Rham and L.~Heisenberg,
  JCAP {\bf 1105}, 025 (2011)
  [arXiv:1104.0155 [hep-th]].
  
\bibitem{Hassan:2011tf} 
  S.~F.~Hassan, R.~A.~Rosen and A.~Schmidt-May,
  JHEP {\bf 1202}, 026 (2012)
  [arXiv:1109.3230 [hep-th]].
  
\bibitem{Hassan:2011zd} 
  S.~F.~Hassan and R.~A.~Rosen,
  JHEP {\bf 1202}, 126 (2012)
  [arXiv:1109.3515 [hep-th]].

\bibitem{Gabadadze:2012tr} 
  G.~Gabadadze, K.~Hinterbichler, J.~Khoury, D.~Pirtskhalava and M.~Trodden,
  Phys.\ Rev.\ D {\bf 86}, 124004 (2012)
  [arXiv:1208.5773 [hep-th]].


\bibitem{Gumrukcuoglu:2011ew} 
  A.~E.~Gumrukcuoglu, C.~Lin and S.~Mukohyama,
  JCAP {\bf 1111}, 030 (2011)
  [arXiv:1109.3845 [hep-th]].

\bibitem{Gumrukcuoglu:2011zh} 
  A.~E.~Gumrukcuoglu, C.~Lin and S.~Mukohyama,
  JCAP {\bf 1203}, 006 (2012)
  [arXiv:1111.4107 [hep-th]].

\bibitem{Gumrukcuoglu:2012aa} 
  A.~E.~Gumrukcuoglu, C.~Lin and S.~Mukohyama,
  Phys.\ Lett.\ B {\bf 717}, 295 (2012)
  [arXiv:1206.2723 [hep-th]].

\bibitem{DeFelice:2013awa} 
  A.~De Felice, A.~E.~Gumrukcuoglu, C.~Lin and S.~Mukohyama,
  arXiv:1303.4154 [hep-th].

\bibitem{D'Amico:2012zv} 
  G.~D'Amico, G.~Gabadadze, L.~Hui and D.~Pirtskhalava,
  arXiv:1206.4253 [hep-th].

\bibitem{Haghani:2013eya} 
  Z.~Haghani, H.~R.~Sepangi and S.~Shahidi,
  arXiv:1303.2843 [gr-qc].
  
\bibitem{Gumrukcuoglu:2013nza} 
  A.~E.~Gumrukcuoglu, K.~Hinterbichler, C.~Lin, S.~Mukohyama and M.~Trodden,
  arXiv:1304.0449 [hep-th].
  
\bibitem{D'Amico:2013kya} 
  G.~D'Amico, G.~Gabadadze, L.~Hui and D.~Pirtskhalava,
  arXiv:1304.0723 [hep-th].
  
\bibitem{D'Amico:2011jj} 
  G.~D'Amico, C.~de Rham, S.~Dubovsky, G.~Gabadadze, D.~Pirtskhalava and A.~J.~Tolley,
  Phys.\ Rev.\ D {\bf 84}, 124046 (2011)
  [arXiv:1108.5231 [hep-th]].
  
\bibitem{Huang:2012pe} 
  Q.~-G.~Huang, Y.~-S.~Piao and S.~-Y.~Zhou,
  Phys.\ Rev.\ D {\bf 86}, 124014 (2012)
  [arXiv:1206.5678 [hep-th]].
  
\bibitem{DeFelice:2013bxa} 
  A.~De Felice, A.~E.~Gumrukcuoglu, C.~Lin and S.~Mukohyama,
  arXiv:1304.0484 [hep-th].
  
  
\bibitem{Andrews:2013ora} 
  M.~Andrews, G.~Goon, K.~Hinterbichler, J.~Stokes and M.~Trodden,
  arXiv:1303.1177 [hep-th].
  
\bibitem{Kluson:2013hoa} 
  J.~Kluson,
  arXiv:1305.6751 [hep-th].


\bibitem{Hinterbichler:2013dv} 
  K.~Hinterbichler, J.~Stokes and M.~Trodden,
  arXiv:1301.4993 [astro-ph.CO].
  
\bibitem{Goon:2010xh} 
  G.~L.~Goon, K.~Hinterbichler and M.~Trodden,
  Phys.\ Rev.\ D {\bf 83}, 085015 (2011)
  [arXiv:1008.4580 [hep-th]].

\bibitem{Gratia:2012wt} 
  P.~Gratia, W.~Hu and M.~Wyman,
  Phys.\ Rev.\ D {\bf 86}, 061504 (2012)
  [arXiv:1205.4241 [hep-th]].

\bibitem{Wyman:2012iw} 
  M.~Wyman, W.~Hu and P.~Gratia,
  Phys.\ Rev.\ D {\bf 87}, 084046 (2013)
  [arXiv:1211.4576 [hep-th]].

\bibitem{Comelli:2012db} 
  D.~Comelli, M.~Crisostomi and L.~Pilo,
  JHEP {\bf 1206}, 085 (2012)
  [arXiv:1202.1986 [hep-th]].

\bibitem{Koyama:2011wx} 
  K.~Koyama, G.~Niz and G.~Tasinato,
  JHEP {\bf 1112}, 065 (2011)
  [arXiv:1110.2618 [hep-th]].

\end{thebibliography}
\end{document}